# Interplay between structure and magnetism in the low-dimensional spin system: K(C$_8$H$_{16}$O$_4$)$_2$CuCl$_3$·H$_2$O

*Natalija van Well[a,d*], Michael Bolte[b], Bernard Delley[c], Bernd Wolf[a], Michael Lang[a], Jürg Schefer[d], Christian Rüegg[d,e], Wolf Assmus[a], and Cornelius Krellner[a]*

[a]*Physikalisches Institut, Goethe-Universität Frankfurt, D-60438 Frankfurt am Main, Germany*
[b]*Institut für Anorganische und Analytische Chemie, Goethe-Universität Frankfurt, D-60438 Frankfurt am Main, Germany*
[c]*Condensed Matter Theory Group, Paul Scherrer Institute, CH-5232 Villigen, Switzerland*
[d]*Laboratory for Neutron Scattering and Imaging, Paul Scherrer Institute, CH-5232 Villigen, Switzerland*
[e]*Department for Quantum Matter Physics, University of Geneva, CH-1211 Geneva, Switzerland*

*Materials based on a crown ether complex together with magnetic ions, especially Cu(II), can be used to synthesize new low dimensional quantum spin systems. We have prepared the new crown ether complex Di-μ-chloro-bis(12-crown-4)-aquadichloro-copper(II)-potassium, K(C$_8$H$_{16}$O$_4$)$_2$CuCl$_3$·H$_2$O (1), determined its structure, and analyzed its magnetic properties. Complex (1) has a monoclinic structure and crystallizes in space group P2$_1$/n with the lattice parameters of a = 9.5976(5) Å, b = 11.9814(9) Å, c = 21.8713(11) Å and β = 100.945(4)°. The magnetic properties of this compound have been investigated in the temperature range 1.8 K – 300 K. The magnetic susceptibility shows a maximum at 23 K, but no 3-D long range magnetic order down to 1.8 K. The S=1/2 Cu(II) ions form antiferromagnetically coupled dimers with Cu–Cl distances of 2.2554(8) Å and 4.683(6) Å, and a Cu–Cl–Cu angle of 115.12(2)° with 2J$_{dimer}$= -2.96 meV (-23.78 cm$^{-1}$). The influence of H$_2$O on the Cl-Cu-Cl exchange path is analyzed. Our results show that the values of the singlet-triplet splitting are increasing considering H$_2$O molecules in the bridging interaction. This is supported by Density functional theory (DFT) calculations of coupling constants with Perdew and Wang (PWC), Perdew, Burke and Ernzenrhof (PBE) and strongly constrained and appropriately normed (SCAN) exchange-correlation function show excellent agreement for the studied compound.*

## Introduction

The synthesis with flexible molecules like crown ether[1] in low dimensional systems is of great interest to study the modification of physical properties depending on the structural variations of such materials. The syntheses lead to additional dimerized and chain compounds and to the different low dimensional variation such as clusters in the metal organic material class. It is notable that in such systems a different magnetic behavior ranging from ferromagnetic to antiferromagnetic interactions has been found[2-5,6,7]. The compounds with two halo bridges have been extensively studied, as the structural variety is very large, and the existence of a correlation between their magnetic behavior and the structural parameters is less obvious than for other bridged compounds. However, it is difficult to establish a simple magneto-structural relationship between the strengths of the magnetic exchange coupling constant J and the Cu-Cl-Cu bridging angle, the Cu-Cl bridging bonds or the Cu-Cu distances[8-14,15]. It is also worth mentioning that the nature of the magnetic interaction depends on the orbital overlapping (characterized by the bonding distances and the angles). The metal atoms in dihalo-bridged Cu(II) complexes are usually four- or five-coordinated with different types of terminal ligands, and show as a result a wide range of values for the coupling constant. In small clusters of two magnetic ions, embedded in a non-magnetic ligand group, the copper ions interact mainly via the super-exchange Heisenberg interaction. Several theoretical and experimental analyses have been carried out to clarify the structures and magnetic properties of the chloro-bridges Cu(II) dimers. Hatfield et al. studied dihalo-bridged Cu(II) dinuclear compounds and realized that the experimental coupling constant is correlated with the ϕ/R ratio, whereby ϕ is the Cu-Cl-Cu angle and R the longest Cu-Cl distance[15]. Although previous theoretical studies for this Cu(II) dinuclear compounds have been reported, there are still a wide range of aspects to be considered due to the large structural variety of this kind of material[16-19]. The most important factors, which have an influence on the coupling mechanism, are the connecting route of both copper atoms, the structure variations in the bridging region and nature of the terminal ligands[12]. The aim of applying a DFT calculation is, to examine the exchange coupling phenomenon for this compound (1) to receive a reference value of the energy differences between states, which can be compared with the experimental one. The coupling constants for a system with two unpaired electrons comply with J= $E_S$ - $E_T$, where $E_S$ and $E_T$ are the energies of the singlet and the triplet states, respectively. To calculate the difference between the energy states, the broken symmetry (BS) approach[20] is used, which consists of executing unrestricted calculations for low spin molecular systems.

In this article we focus on the investigation of the structural and magnetic properties of a newly synthesized low-dimensional spin system. Therefore, the detailed investigation of materials of the same type, like the new (1), is of great interest, as it significantly contributes to a better understanding of the magnetic properties of such coordinating compounds. After having described the experimental details, we present the results of the crystal growth and the structure characterization. Then, we provide experimental details of the physical properties for the low-dimensional spin system (1), followed by the conclusion and outlook.

## Experimental details

For the crystal structure determination a STOE IPDS-II diffractometer with a Genix microfocus tube and with mirror optics was used. All crystals were measured at 173 K. The data were scaled using the frame-scaling procedure in the X-AREA program system[21]. The structures were solved by direct methods and refined with full-matrix least-squares techniques on $F^2$ using the program SHELXL[22]. The H atoms bonded to O in (1) were refined with the O-H distances restrained to 0.84(2) Å. The crystallographic data can be obtained from the ESI†. The experiments to study the magnetic properties on the single crystal of this compound were performed by using Quantum Design PPMS. The measurement was performed in a magnetic field at 1 T and a temperature range from 1.8 K to 300 K. The sample of (1) was not oriented in the magnetic field. The magnetic moment from the sample holder with varnish, which was used to fix the sample, is around -9x10$^{-6}$ emu. The EPR spectra of the powder sample were recorded with a spectrometer Bruker ESR in X-band. The measurement was performed with a powder sample of (1) at 200 K in a magnetic field from 2400 G to 3800 G.

## Crystal growth and structure determination

Single crystals of compound (1) were grown from solution using the evaporation method. For the compound (1) the crystalline reagents KCl (Suprapur, Merk), LiCl (99.9%, ChemPur), $CuCl_2 \cdot H_2O$ (Analar Normapur, Merck) and 12-crown-4 (98%, Merk) in a 1:1:1:1 molar ratio were used. The crystalline reagents were dissolved in Acetonitril (99.98%, Roth). LiCl served as an additive. The single crystals were grown at the bottom of a vessel over a period of about five to six months at around 4°C. A typical example for such crystals is shown in Figure 1. These crystals have a size between 1 and 10 mm. The material is not stable at room temperature, and, therefore, the preparation is difficult.

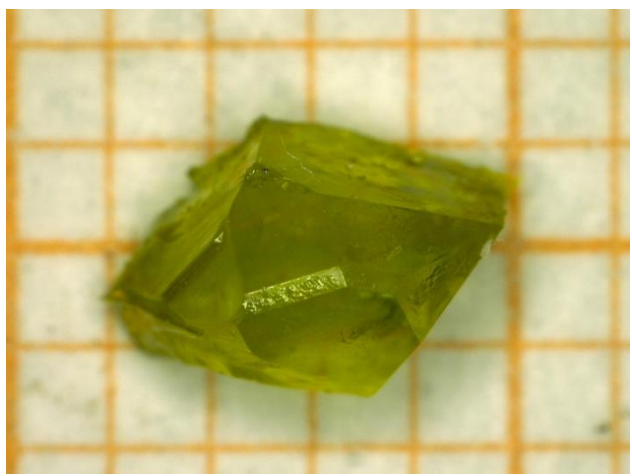

**Figure 1:** Example of crystal of $K(C_8H_{16}O_4)_2CuCl_3 \cdot H_2O$

During the preliminary work for the crystal growth of the Cu based metal organic compounds with different crown ether molecules not only new compositions were found, but also the influence of different solvents regarding crystal growth was investigated[23].

The crystal structure of (1) was determined by single crystal X-ray diffraction. This metal-organic compound is monoclinic and crystallized in the space group $P2_1/n$. A perspective view of (1) is shown in Figure 2. The newly crystallized metal organic compound (1) shows two copper atoms in this dimer, which are 5.9988(8) Å apart. In Figure 2 the interaction in the dimer compound through $Cu_2Cl_2$ bondings is shown. The bondings are visualized by dashed lines.

The asymmetric unit of (1) as shown in Figure 3a consists of two crown ether rings, which are connected through four oxygen atoms to the K$^+$ ion. The K...O distances do not differ significantly. They range from 2.818(2) Å for K1-O1 to 3.062(2) Å for K1-O2. The K$^+$ ion is additionally linked (K1...Cl1 3.2526(10) Å) to a Cl ligand of a $CuCl_3H_2O$ moiety. The Cu centre features a distorted tetrahedral coordination with bond angles ranging from 92.65(7)° for O1W-Cu1-Cl3 to 140.23(8)° for O1W-Cu1-Cl2. The Cu-Cl distances are in the same range: 2.1906(8) Å (Cu1-Cl2), 2.2165(8) Å (Cu1-Cl1) and 2.2554(8) Å (Cu1-Cl3). The water molecule bonded to Cu (Cu1-O1W 1.979(2) Å) forms an O-H...O hydrogen bond to one O atom of the crown ether ring. Two $K(C_8H_{16}O_4)_2CuCl_3 \cdot H_2O$ units are connected by a $Cu_2Cl_2$ bridging unit (see Figure 3b, dashed line) to form a centrosymmetric dimer. The shortest Cu...Cu distances are found along the crystallographic b-axis. Between two bonded $CuCl_3H_2O$ moieties the Cu...Cu distance amounts to 5.9988(8) Å. The shortest Cu...Cu distance between two $CuCl_3H_2O$ moieties, which are not bonded, is also in direction of the b axis and has a value of 6.7532(8) Å.

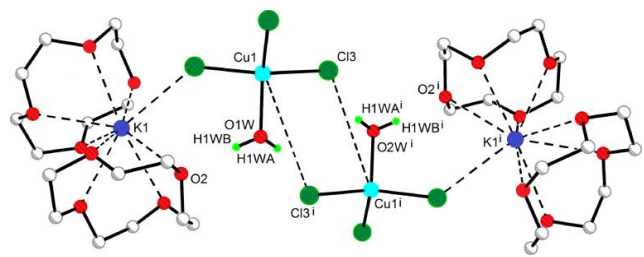

**Figure 2:** Perspective view of a dimer of $K(C_8H_{16}O_4)_2CuCl_3 \cdot H_2O$, symmetry code: (i) –x, -y, -z

The crystal packing is characterized by layers of $K(C_8H_{16}O_4)_2$ units arranged parallel to the (0 0 1) plane. In between these planes layers of $CuCl_3H_2O$ moieties are located. Figure 3b presents the partial packing diagram for (1).

Crystal data, data collection and structure refinement details are summarized in Table 1.

## Table 1

**Table 1:** Selected crystallographic data for K(C$_8$H$_{16}$O$_4$)$_2$CuCl$_3$·H$_2$O

| Chemical formula | K(C$_8$H$_{16}$O$_4$)$_2$CuCl$_3$·H$_2$O |
| --- | --- |
| M$_w$ | 579.42 |
| radiation, λ [Å] | MoK$_\alpha$, 0.71073 |
| space group | $P2_1/n$, monoclinic |
| a [Å] | 9.5976(5) |
| b [Å] | 11.9814(9) |
| c [Å] | 21.8713(11) |
| β [°] | 100.945(4) |
| V [Å$^3$] | 2469.3(3) |
| Z | 4 |
| ρ$_{calc}$ [mg/cm$^{-3}$] | 1.559 |
| absorption coefficient | 1.419 |
| crystal size [mm] | 0.21×0.15×0.04 |
| total reflns (R$_{int}$) | 25729(0.0814) |
| reflections/restraints/parameters | 5712/2/279 |
| GOF on F$^2$ | 1.061 |
| R [F$^2$ > 2σ (F$^2$)] | 0.0515 |
| wR (F$^2$) | 0.0575 |
| largest electron density difference [e Å$^{-3}$] | 1.579, -0.589 |

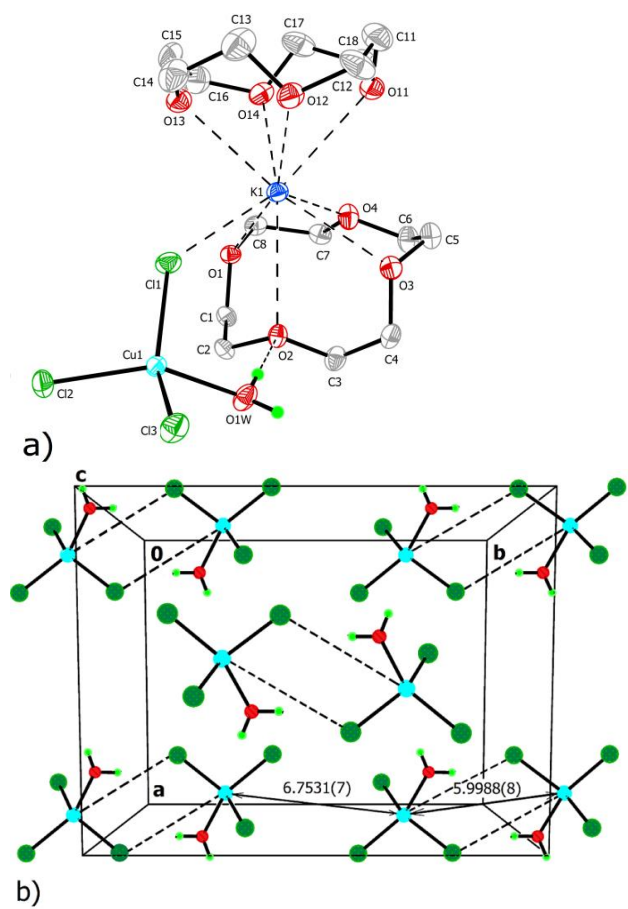

**Figure 3:** a) Perspective view of the asymmetric unit of (1), showing the atom numbering scheme and the thermal ellipsoids, which are drawn at a 50% probability level, b) Partial packing diagram for (1), K cations and crown ether molecules omitted for clarity. The Cu dimer units have a distance of 5.9988(8) Å and 6.7531(7) Å between the dimers

## Physical properties

Magnetic susceptibility measurements show that the magnetic properties of (1) are dominated by a dimer interaction between neighboring Cu(II) ions. As presented in Figure 4a the maximum of their susceptibility is at around 23 K. Towards lower temperatures the drop of the susceptibility shows the antiferromagnetic behavior. No 3D long range magnetic order was observed down to 1.8 K.

All results are consistent with dimer units and with a gapped ground state. With respect to the dimer model, if the magnetic susceptibility decreases with the temperature dropping to $T = 0$, the ground state is $S = 0$.

A small extrinsic paramagnetic impurity contribution (4.4 % of the total number of Cu spins, see Figure 4b), the temperature-independent diamagnetic term $\chi_{dia}$=-0.002962·10$^{-6}$m$^3$·mol$^{-1}$, which is estimated from Pascal`s constants, and the background contribution of the sample holder, were taken into account with respect to the data in Figure 4a[24]. The Curie-Weiss temperature $\Theta$ and the Curie constant $C$ can be extracted from the function of the inverse susceptibility in the temperature range 100 K < $T$ < 300 K (see Figure 4c). Both were calculated resulting to $C = (5.652\pm0.001)10^{-6}$ Km$^3$·mol$^{-1}$ and $\Theta = (-12.8\pm0.4)$ K. In order to describe the interaction between the two copper ions inside a dimer, a Heisenberg hamiltonian has been used:

$$\hat{H}_{ex} = -2J \sum_{i<j} \hat{S}_i \cdot \hat{S}_j \qquad (I)$$

Thus the magnetic data has been fitted using the Bleaney and Bowers equation[25] for Cu(II) dimers. For calculations of the total susceptibility the following formula was used:

$$X_{mol} = (1-x)\frac{\mu_0 N_A \mu_B^2}{3k_B T} g^2 \left[1 + \frac{1}{3}exp\left(\frac{-2J}{k_B T}\right)\right]^{-1} + x\frac{C_{mol}}{T} + X_0 \quad (II)$$

The *g*-factor was obtained from the EPR measurement and has the value $g = 2.074\pm0.01$. The temperature-independent paramagnetic part ($\chi_0$) amounted to $8\cdot10^{-10}$m$^3$·mol$^{-1}$.

The magnetic exchange coupling determined from the fit of the susceptibility measurements in Figure 4a gives a value of $2J_{dimer}$= -2.96 meV (-23.78 cm$^{-1}$).

Furthermore, the sum of the squares of the deviations from the calculated curve, SD = $\sum_{i=1}^{n}((X_{i\,exp} - X_{i\,mol})/X_{i\,exp})^2$, is 3.9·10$^{-2}$. The result is a network of two magnetic Cu(II) ions as shown in Figure 4d. The structural information of this new material (1) results in the schematic depiction as shown in Figure 4d. The geometry of the Cu dimer is described as a two distorted tetrahedral coordination. Inside the Cu$_2$Cl$_2$ unit the Cu-Cu distance is 5.9988(8) Å, and the Cu-Cl-Cu angle equals to 115.12(2)°. The Cu dimer unit shows the longest distance between the metal atoms, according to literature at the time. Nevertheless, the moderate interaction between Cu-Cu in the dimer led to the consideration and verification of an alternative next neighbor exchange coupling in the dimer[23]. Figure 4d shows the structural information of compound (1) as a schematic depiction. The theoretical calculation helps to understand the influence of the H$_2$O molecule on the Cu$_2$Cl$_2$ bridging, and will clarify the role of H$_2$O in the interaction description of the investigated compound.

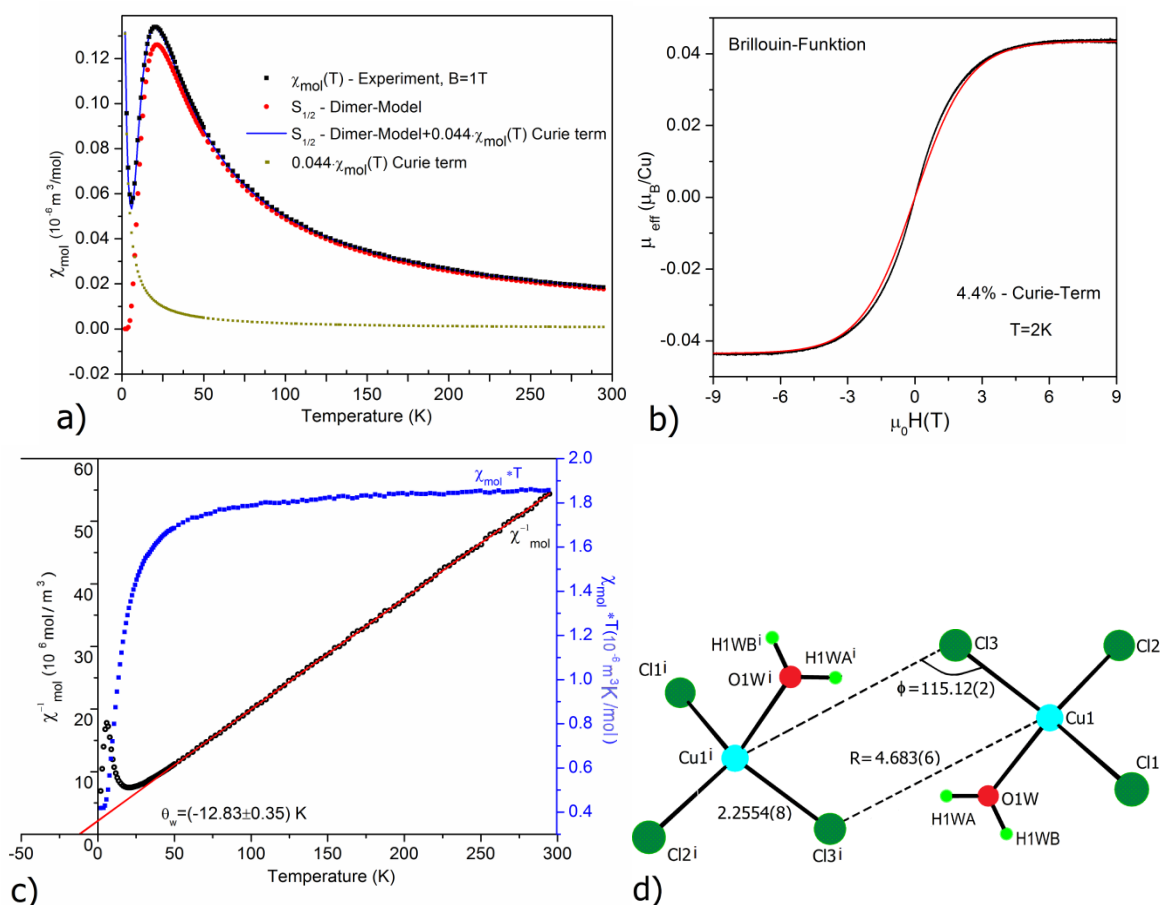

**Figure 4:** a) Susceptibility of (1) and fit using a dimer model ($S1 = S2 = ½$), the solid line represents the calculated curve, according to Bleaney-Bowers equation, b) Magnetization at 2 K and fit using Brillouin-function, c) $\chi_{mol}(T)*T$ vs. $T$ and inverse susceptibility $\chi^{-1}_{mol}(T)$ depending on temperature, d) Schematic depiction of the Cu(II) dimer with its coordination, $K$ cations and crown ether molecules omitted for clarity; the atoms with suffix (i) were generated by the symmetry operator -x, -y, -z

**Table 2:** Selected structural and magnetic properties of chloro-bridging Cu(II) dimers

| Compound | Cu1-Cu1$^i$ (Å) | Cu1-Cl3$^i$ (R,Å) | Cu1-Cl3-Cu1$^i$ ($\phi$,°) | $\phi/R$ (°/Å) | $2J$ (cm$^{-1}$) | Ref. |
|---|---|---|---|---|---|---|
| 1. K(C$_8$H$_{16}$O$_4$)$_2$CuCl$_3$·H$_2$O | 5.9988(8) | 4.683(6) | 115.12(2) | 24.25 | -22.7 | this work |
| 2. [{Cu(MeL)Cl}$_2$][ClO$_4$]$_2$ | 3.891(2) | 2.891(2) | 94.8(7) | 32.79 | -7.78 | 29 |
| 3. [Cu(dmen)Cl$_2$]$_2$ | 3.458(3) | 2.734(3) | 86.13(8) | 31.5 | -2.1 | 28 |
| 4. [Cu(4-Metz)(DMF)Cl$_2$]$_2$ | 3.721 | 2.724 | 95.3 | 34.99 | -3.6 | 28 |
| 5. [Cu(tmen)Cl$_2$]$_2$ | 4.089(4) | 3.147(4) | 96.8(1) | 30.76 | -5.6 | 30 |
| 6. [Cu(2-pic)$_2$Cl$_2$]$_2$ | 4.404(1) | 3.364(1) | 100.63(3) | 29.91 | -7.4 | 8 |
| 7. [Cu(tmso)Cl$_2$]$_2$ | 3.737(2) | 3.020(2) | 88.5(1) | 29.3 | -16 | 31 |
| 8. [Cu(terpy)Cl]$_2$[PF$_6$]$_2$ | 3.510(1) | 2.720 | 89.9 | 33.1 | -5.8 | 32 |
| 9. [Cu(GuaH)Cl$_3$]$_2$ 2H$_2$O | 3.575 | 2.447(1) | 97.9 | 40.01 | -82.6 | 29, 33 |
| 10. [Cu(Et$_3$en)Cl$_2$]$_2$ | 3.703(1) | 2.728(1) | 94.84(4) | 34.75 | +0.1 | 28 |
| 11. [Cu(dmgH)Cl$_2$]$_2$ | 3.445 | 2.698 | 88.0 | 32.62 | +6.3 | 8 |

**Abbreviations:** dmgH, dimethylglyoxime; dmen, N,N-dimethylenediamine; GuaH, guaninium; Et$_3$en, N,N, N'-triethylethylenediamine; 4-Metz, 4-methylthiazole; N,N-dimethylformamide; tmso, tetramethylene sulfoxide; tmen, N,N, N0, N0-tetramethylethylenediamine; 2-pic, 2-methylpyridine; MeL = methyl[2-(2-pyridyl)ethyl](2-pyridylmethyl)amine; terpy = N,N',N'-terpyridine; DMF = N,N-dimethylformamide

**Table 3:** Experimental value of coupling constants and calculated energy difference $\Delta E$ with different methods for dinuclear Cu(II) complexes.

| Compound | $2J_{exp}$ (cm$^{-1}$)/ Ref | Model | $\Delta E = E_{BS} - E_T$ (cm$^{-1}$) | | |
|---|---|---|---|---|---|
| | | | PWC | PBE | SCAN |
| K(C$_8$H$_{16}$O$_4$)$_2$CuCl$_3$ | - | dimer | -15 | -12 | -8 |
| K(C$_8$H$_{16}$O$_4$)$_2$CuCl$_3$·H$_2$O | -23.78 [this work] | dimer incl. H$_2$O | -125 | -83 | -31 |
| K(C$_8$H$_{16}$O$_4$)$_2$CuCl$_3$·H$_2$O | -23.78 [this work] | crystal / per dimer | -124 | -55 | -19 |
| [Cu(tmso)Cl$_2$]$_2$ | -16 [31] | dimer | -63 | -40 | -17 |
| [Cu(GuaH)Cl$_3$]$_2$ 2H$_2$O | -82.6 [29, 33] | dimer | -226 | -137 | -32 |

**Abbreviations:** GuaH, guaninium; tmso, tetramethylene sulfoxide. For the known compounds the structure parameter from the CCDC database (measured at room temperature) was used. The structure optimization was performed with the local density functional DMol$^3$ code.

The bridging ligands between the metal ions influence the value and the magnitude of the exchange coupling constant, which are subject to the various types of overlap interactions between the metal d-orbitals and the ligand orbitals. Magneto-structural correlations have been identified for certain types of binuclear Cu(II) complexes[8-13,26,27]. A summary of a bulk of known magnetic data on salts containing asymmetrical Cu–Cl bridges can be found in literature[8,10,11,28]. A lot of factors have been analyzed through different calculations, to understand the influence of such factors on the coupling constants in such bridged compounds. One of the most important factors is the specific path of connecting both copper atoms. In addition, also the structural variations of distances and angles in the bridging region, and the nature of the different ligands, being involved in bridging, are of great importance[12]. Table 2 compares the new compound (1) with other dimers described in literature. The comparison is related to the above mentioned three important factors.

With regard to the investigation of complex (1), the exchange coupling constant $2J$ depends on the magnitude of the angle of the bridge Cu1–Cl3–Cu1$^i$ ($\phi$) as well as on the bond length of the longest, out of plane bond ($R$) for the compounds, which have defining basal planes. In this case, ($R$) is the Cu1-Cu3$^i$ bond (see Figure 4d). A singlet ground state with antiferromagnetic character is found for values of the quotient $\phi/R$ lower than 32.6°/Å and higher than 35.8°/Å, whereas a triplet ground state with ferromagnetic character appears, when the quotient $\phi/R$ is between the two fore-mentioned values[15], as shown in Figure 5. Thus, the result for the exchange coupling constant for complex (1) is validated by the already known data from literature.

In order to provide a more detailed analysis of the Cu-Cu interactions for complex (1), and as we like to compare the magnetic properties with the density functional theory (DFT) results, DFT calculations for the structure, as determined by the single crystal X-ray diffraction refinements at 173 K (for the investigated compound), were done by applying the generalized gradient approximation (GGA's) for the exchange-correlation energy and the local spin density (LSD) description exchange and correlation functional to determine the correlation between the Cu(II) 3d orbitals and ligands orbitals. Within the local density approximation (LDA) approach, we consider exchange-correlation functional of Perdew and Wang (PWC)[34]. Our generalized gradient approximation (GGA) is based on Perdew, Burke and Ernzenrhof (PBE)[35] and the strongly constrained and appropriately normed (SCAN)[36,37] exchange-correlation function. The DFT calculations were done by using the DMol$^3$ code with its standard DNP basis set[38]. Furthermore, the calculations for the crystal model were done with DMol$^3$ [39], using 2x2x2 k-points mesh.

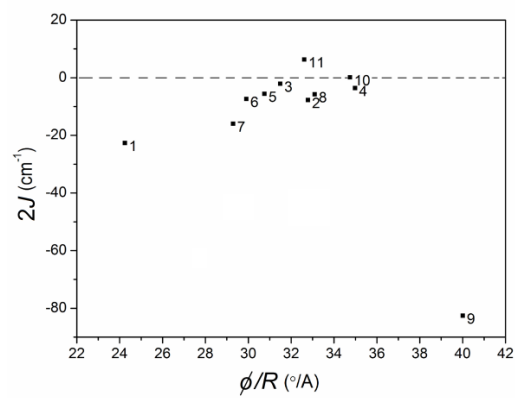

**Figure 5:** Plot of the singlet–triplet splitting $2J$ as a function of $\phi/R$ in different binuclear copper complexes. Details of compounds numbered 1 to 11 are listed in Table2.

In the case of a system with two weakly coupled spins 1/2, spin-unrestricted DFT allows to look for a ground state with two unpaired spins, which can be interpreted as the triplet state with energy $E_T$. Alternatively, DFT allows searching for its lowest energy state with Energy $E_{BS}$. In the present, case spin-unrestricted and spin-restricted DFT agree fully concerning this $E_{BS}$ state. By the virtue of the functional, this state is an approximation to the ground state, which is a singlet state in the present case of the investigated compounds. By consequence, the energy difference can be interpreted as $2J = E_{BS} - E_T$ of the Heisenberg model for combined maximum spin $S = 1$. This view has been challenged by Noodleman[20,40]. His argument can be briefly recapitulated. While the determinant related to $E_T$ clearly is a representative of the triplet state, the one for $E_{BS}$ is not closely representing a pure spin state. Rather, it is a fifty-fifty mixture of the triplet and the singlet state. By consequence $J = E_{BS} - E_T$, half of the full singlet-triplet splitting.

The results of the energy difference calculated with different functionals in comparison to the new compound (1) and the two already known dimer compounds can be seen in Table 3. We realized for (1) that the negative charge of $CuCl_3H_2O$ favors a wave function delocalization, thus resulting in increasing exchange interaction between Cu-Cu ions. The calculations with different methods of the investigated compounds with and without a $H_2O$ show an overall decrease of the singlet-triplet splitting. A comparison of the calculated and experimental values shows that the $H_2O$ molecule favours the interaction in the $Cu_2Cl_2$ bridging. The calculated value of the energy difference $\Delta E = -31 cm^{-1}$ using the SCAN exchange-correlation function shows an excellent correlation with the experiment value ($2J_{exp}=-23.78 cm^{-1}$). The deviation from the optimal geometry of the individual molecules is due to packing forces that are common using the experimental magnetic susceptibility from solid samples to obtain the coupling constant.

The coupling constant of $[Cu(tmso)Cl_2]_2$ was also calculated with the different functionals and show the same trend of the value as (1). The value of $\Delta E=-17 cm^{-1}$ is in excellent agreement with the experimental value for this compound ($2J_{exp}=-16 cm^{-1}$).

For the last compound in Table 3 ($[Cu(GuaH)Cl_3]_2 \cdot 2H_2O$) the theoretical calculation, already known from literature, results in $\Delta E=-60$ ($cm^{-1}$), which was performed with DFT based of B3LYP method[12]. All calculated values of the energy difference for this compound in this work show slightly increasing differences to the experimental values. Nevertheless, the trend is the same like that for the other two compounds. To sum up, the SCAN method seems to be the best one of the used functionals for the theoretical approximation of the singlet-triplet splitting of the investigated compounds.

## Conclusion and outlook

With the synthesis of flexible crown ether molecules, it has succeeded to grow the compound $K(C_8H_{16}O_4)_2CuCl_3 \cdot H_2O$ (1), which belong to the low-dimensional spin systems. The new material (1) has a monoclinic structure and crystallized in the space group $P2_1/n$. The magnetic ground state of (1) was determined by magnetic measurements and found to be a non-ordered material with dinuclear Cu(II) units, which are antiferromagnetically coupled with an intradimer coupling $2J_{dimer}= -2.96$ meV ($-23.78$ cm$^{-1}$). The dinuclear units $Cu_2Cl_2$ with the Cu–Cl distances of 2.2554(8) Å and 4.683(6) Å, and a Cu–Cl–Cu angle of 115.12(2)° in this compound were formed. The $H_2O$ molecule plays an important role in the interaction of the $Cu_2Cl_2$ bridging, and is responsible for the increasing exchange interaction between Cu-Cu ions. The energy difference calculation with SCAN ($\Delta E=-31$ cm$^{-1}$) shows an excellent correlation with the experimental value ($2J_{exp}=-23.78$ cm$^{-1}$).

Then, for example, such investigation of (1) will provide detailed information about the correlations and interactions in this new material, as it is likely dominated by the quantum dimer units. Extensive knowledge about the relation between the structure of the new composition and its magnetic properties can help to comprehend the variation of magnetic properties in case of modification of this compound by way of a substitution with Br. For future material development of new quantum spin systems, metal organic materials are prioritized, whereby transition metal ions like Cu(II) with partly filled d-shell or stable organic radicals are used as flexible groups. By using organic molecules, a high degree of flexibility is achieved. This provides for the systematic investigation and the optimization of the interaction of many particle systems by varying the chemical and physical parameter. Furthermore, it is also interesting to study the changes of the structural and physical properties of such materials by using different crown ether molecules for the synthesis or by using a variation of ligands.

In addition, experiments under pressure allow new insights into the structural and physical properties of compounds crystallized with different flexible crown ether molecules, for example, to study under pressure the continuous tuning within the isotropic or isolated chain limits, especially, as even a moderate pressure should be sufficient to vary the distance between atoms, due to the fact that the coupling of the molecules of metalorganic compounds is weaker than those of inorganic materials. Hereby, pressure acts as control parameter to cause magnetic interaction between the involved magnetic ions, even if such magnetic ions haven't shown an interaction before. Such phenomena are also appropriate for theoretical modelling.

## Acknowledgements

The authors thank K. Krämer from the Department of Chemistry and Biochemistry, University of Bern for fruitful discussions and S. Waldschmidt for her work on this material during her master theses. This work was supported by the Paul Scherrer Institut, Villigen, Switzerland, and the Deutsche Forschungsgemeinschaft through SFB/TRR49 and the research fellowship for the project – WE-5803/1-1.

## References

1   C. J. Pedersen, H. K. Frensdorff, Angew. Chem., 1972, 84, 16
2   A. Escuer, G. Aromi, Eur. J. Inorg. Chem., 2006, 4721
3   G. Aromi, E. K. Brechin, R. Winpenny, Eds., Springer-Verlag Berlin, Berlin 2006, Vol. 122, pp 1-67

4 N. Motokawa, H. Miyasaka, M. Yamashita, K. R. Dunbar, Angew. Chem. Int. Ed., 2008, 47, 7760
5 A. Mishra, A. J. Tasiopoulos, W. Wernsdorfer, E. E. Moushi, B. Moulton, M. J. Zaworotko, K. A. Abboud, G. Christou, Inorg. Chem., 2008, 47, 4832
6 E. Escriva, J. Server-Carrio, J. Garcia-Lozano, J. V. Folgado, F. Sapina, L. Lezama, Inorg. Chim. Acta, 1998, 279, 58
7 Y. Yamada, N. Ueyama, T. Okamura, W. Mori, A. Nakamura, Inorg. Chim. Acta, 1998, 276, 43
8 W. E. Marsch, W. E. Hatfield, D. J. Hodgson, Inorg. Chem., 1982, 21, 2679
9 S. K. Hoffmann, D. K. Towle, W. E. Hatfield, K. Wieghardt, P. Chaudhuri, Mol. Cryst. Liq. Cryst., 1984, 107, 161
10 M. Rodriguez, A. Llobet, M. Corbella, A. E. Martell, J. Reibenspies, Inorg. Chem., 1999, 38, 2328
11 M. Rodriguez, A. Llobet, M. Corbella, Polyhedron, 2000, 19, 2483
12 A. Rodriguez-Fortea, P. Alemany, S. Alvarez, E. Ruiz, Inorg. Chem., 2002, 41, 3769
13 A. M. Schuitema, A. F. Stassen, W. L. Driessen, J. Reedijk, Inorg. Chim. Acta, 2002, 337, 48
14 Y.-H. Chung, H.-H. Wie, Y.-H. Liu, G.-H. Lee, Y. Wang, J. Chem. Soc. Dalton Trans., 1997, 2825
15 W. E. Hatfield, In Magneto-Structural Correlations in Exchange Coupled Systems; Willett, R. D.; Gatteschi, D.; Kahn, O., Eds.; Reidel: Dordrecht, Netherlands, 1985
16 P. J. Hay, J. C. Thibeault, R. Hoffmann, J. Am. Chem. Soc., 1975, 97, 4884
17 A. Bencini, D. Gatteschi, J. Am. Chem. Soc., 1986, 108, 5763
18 O. Castell, J. Miralles, R. Caballol, Chem. Phys., 1994, 179, 377
19 A. Bonamartini-Corradi, L. P. Battaglia, J. Rubenacker, R. D. Willett, T. E. Grigereit, P. Zhou, J. E. Drumheller, Inorg. Chem., 1992, 31, 3859
20 L. Noodleman, J. Chem. Phys., 1981, 74, 5737
21 Stoe & Cie, *X-AREA*, Stoe & Cie, Darmstadt, Germany, 2002
22 G. M. Sheldrick, Acta Crystallogr. Sect., 2008, A64, 112
23 N. van Well, , Innovative und interdisziplinäre Kristallzüchtung, Springer Spektrum, Wiesbaden, 2016,
24 O. Kahn, , Molecular Magnetism, VCV Publishers, New York, 1993
25 B. Bleaney, K. D. Bowers, Proc. R. Chem. London, 1952, A214, 451
26 C. J. Calzado, J. Cabrero, J. P. Malrieu, R. Caballol, J. Chem. Phys., 2002, 116, 3985
27 E. Kavlakoglu, A. Elmali, Y. Elerman, I. Svoboda, Polyhedron, 2002, 21, 1539
28 W. E. Marsh, K. C. Patel, W. E. Hatfield, D. Hodgson, J. Inorg. Chem., 1983, 22, 511
29 S. Mandal, F. Lloret, Inorg. Chimi. Acta, 2009, 362, 27
30 E. D. Estes, W. E. Estes, W. E. Hartfield, and D. J. Hodgson, Inorg. Chem., 1975, 14, 106
31 D. D. Swank, G. F. Needham, R. D. Willett, Inorg. Chem., 1979, 18, 762
32 T. Rojo, J. Darriet, J. M. Dance, D. Beltrán-Porter, Inorg. Chim. Acta, 1982, 64, L105
33 R. F. Drake, H. Van Crawford, N. W. Laney, W. E. Hatfield, Inorg. Chem., 1974, 13, 1246
34 J. P. Perdew, K. Burke, and Y. Wang, Phys. Rev. B, 1996, 54, 16533
35 J. P. Perdew, K. Burke, M. Ernzerhof, Phys. Rev. Lett., 1996, 77, 3865
36 J. Sun, A. Ruzsinszky, and J. P. Perdew, Phys. Rev. Lett., 2015, 115, 036402
37 J. P. Perdew, J. Sun, R. M. Martin, and B. Delley, Inter. J. of Quantum Chem., 2016, 116, 847
38 B. Delley, J. Chem. Phys., 1990, 92, 508
39 B. Delley, J. Chem. Phys., 2000, 113, 7756
40 L. Noodleman, E. R. Davidson, Chem. Phys., 1986, 109